\documentclass[prb,amssymb,twocolumn]{revtex4}

\usepackage{graphicx}% Include figure files

\def\vk{{\bf k}}
\def\Tau{\mathcal{T}}

\def\bra#1{\left\langle#1\right|}
\def\ket#1{\left|#1\right\rangle}

\def\abs#1{\left|#1\right|}
\def\kc#1{\left(#1\right)}
\def\kd#1{\left[#1\right]}
\def\ke#1{\left\{#1\right\}}

\def\sgn{{\rm sgn}}
\def\mode{\text{ }{\rm mod}\text{ }}
\def\trace#1{{\rm Tr}\left[#1\right]}

\def\be{\begin{equation}}       \def\ee{\end{equation}}
\def\bea{\begin{eqnarray}}      \def\eea{\end{eqnarray}}
\def\ba{\begin{array} }
\def\ea{\end{array} }
\def\bnum{\begin{enumerate} }
\def\enum{\end{enumerate}}

\def\nn{\nonumber}
\def\pa{\partial}
\def\=>{\Rightarrow}
\def\>{\rightarrow}

\def\V{\downarrow}

\def\eye2{\mathbb{I}}
\def\vk{{\bf k}}
\def\Tau{\mathcal{T}}

\def\vect#1{\kc{\ba{c}#1\ea}}

\begin{document}
\title{Fermi Surface Topological Invariants for Time Reversal Invariant Superconductors}

\author{Xiao-Liang Qi, Taylor L. Hughes and Shou-Cheng Zhang}
\affiliation{Department of Physics, Stanford University, Stanford,
    CA 94305, USA\\
    Stanford Institute for Materials and Energy Sciences, SLAC
    National Accelerator Laboratory, Menlo Park, CA 94025, USA}

\begin{abstract}
A time reversal invariant (TRI) topological superconductor has a
full pairing gap in the bulk and topologically protected gapless
states on the surface or at the edge. In this paper, we show that in
the weak pairing limit, the topological quantum number of a TRI
superconductor can be completely determined by the Fermi surface
properties, and is independent of the electronic structure away from
the Fermi surface. In three dimensions (3D), the integer topological
quantum number in a TRI superconductor is determined by the sign of
the pairing order parameter and the first Chern number of the Berry
phase gauge field on the Fermi surfaces. In two (2D) and one (1D)
dimension, the $Z_2$ topological quantum number of a TRI
superconductor is determined simply by the sign of the pairing order
parameter on the Fermi surfaces. We also obtain a generic and
explicit expression of the $Z_2$ topological invariant in 1D and 2D.
\end{abstract}

\date{\today}
\maketitle

Since the discovery of quantum spin Hall
effect\cite{kane2005A,bernevig2006a,bernevig2006d,koenig2007},
topological insulators (TI) in both two dimensions (2D) and three
dimensions (3D) have generated great interest both theoretically and
experimentally\cite{fu2007a, hsieh2008,zhang2009,xia2009,chen2009}.
Following this development, recent attention has focused on time
reversal invariant (TRI) topological superconductors and
superfluids\cite{qi2009b,roy2008,schnyder2008,kitaev2009}. There is
a direct analogy between superconductors and insulators because the
Bogoliubov-de-Gennes (BdG) Hamiltonian for the quasi-particles of a
superconductor is analogous to the Hamiltonian of a band insulator,
with the superconducting gap corresponding to the band gap of the
insulator. $^3$He-B is an example of such a topological superfluid
state. This TRI state has a full pairing gap in the bulk, and
gapless surface states consisting of a single Majorana
cone\cite{qi2009b,roy2008,schnyder2008,chung2009}. In fact, the BdG
Hamiltonian for $^3$He-B is identical to the model Hamiltonian of a
3D topological insulator proposed by Zhang {\it et
al}\cite{zhang2009}. In 2D, the classification of topological
superconductors is very similar to that of topological insulators.
Time-reversal breaking (TRB) superconductors are classified by an
integer\cite{volovik1988,read2000}, similar to quantum Hall
insulators\cite{thouless1982}, while TRI superconductors are
classified\cite{qi2009b,roy2008,schnyder2008,kitaev2009} by a $Z_2$
invariant in 1D and 2D, but by an integer ($Z$) class  in 3D. The
integer-valued topological invariant in 3D can be written as a
winding number over the entire momentum space\cite{schnyder2008},
while an explicit expression of the $Z_2$ topological invariants in
1D and 2D has not been obtained in the literature.

Despite the similarity of topological insulators and topological
superconductors, there is a key physical difference between them.
Starting from a Fermi liquid ``normal state" with some attractive
interaction, superconductivity is induced at low enough
temperatures due to the Cooper instability of the Fermi surface.
At least right below the transition temperature, Cooper pairing is
only important around the Fermi surface, so the topological
properties of such a superconductor are determined {\it
completely} by the properties in the neighborhood of the Fermi
surface, rather than that of the full Brillouin zone, as in the
case of a topological insulator.

Motivated by such an observation, in this letter we obtain simple
and explicit physical criteria for TRI topological
superconductivity in 1, 2 and 3 dimensions in the {\it weak
pairing limit}, where the pairing is only important in a small
neighborhood of the Fermi surface. In 3D, the Fermi surface
topological invariant (FSTI) of a TRI superconductor is determined
by the sign of the pairing order parameter and the first Chern
number of each Fermi surface. Here the first Chern number of a
Fermi surface is defined by the net flux of the Berry phase gauge
field penetrating the Fermi surface. This is quantized as long as
the Fermi surface is a smooth two-dimensional manifold. Based on
this Fermi surface criterion for the 3D topological
superconductor, the $Z_2$ FSTI's in 1D and 2D can be obtained by
dimensional reduction\cite{qi2008}. The criteria for a $Z_2$
nontrivial superconductor in the weak pairing limit is simple: a
TRI superconductor is nontrivial (trivial) if there are an odd
(even) number of Fermi surfaces with a negative pairing order
parameter. For example, the superconductivity in a 2D Rashba
system is nontrivial if the pairing on the two Fermi surfaces has
opposite sign. Inspired by such Fermi surface formulas, we also
obtain an explicit expression of the $Z_2$ FSTI in 1D and 2D which
applies to generic superconductors beyond the weak pairing limit.

{\bf Topological invariant in 3D TRI superconductors.}  We start
from a generic mean-field Hamiltonian of a 3D TRI superconductor,
which can be written in momentum space as\cite{SOM}
\begin{widetext}
\begin{eqnarray}
H&=&\sum_{\vk}\left[\psi_\vk^\dagger
h_\vk\psi_\vk+\frac12\left(\psi_{\vk}^\dagger\Delta_\vk\psi_{-\vk}^{\dagger
T}+H.c.\right)\right]\equiv\sum_{\vk}\Psi_\vk^\dagger
H_\vk\Psi_\vk\nonumber\\
\text{with~}
\Psi_\vk&=&\frac1{\sqrt{2}}\left(\begin{array}{c}\psi_\vk-i\Tau\psi^\dagger_{-\vk}\\\psi_\vk+i\Tau\psi^\dagger_{-\vk}
\end{array}\right),~H_\vk=\frac12\left(\begin{array}{cc}
&h_\vk+i\Tau\Delta_\vk^\dagger\\h_\vk-i\Tau\Delta_\vk^\dagger&\end{array}\right).\label{HBdG}
\end{eqnarray}
\end{widetext} In general, $\psi_\vk$ is a vector with $N$ components and
$h_\vk$ and $\Delta_\vk$ are $N\times N$ matrices. The matrix
$\Tau$ is the time-reversal matrix satisfying $\Tau^\dagger h_\vk
\Tau=h_{-\vk}^T$, $\Tau^2=-\mathbb{I}$ and $\Tau^\dagger
\Tau=\mathbb{I}$. We have chosen a special basis in which the BdG
Hamiltonian $H_\vk$ has a special off-diagonal form. It should be
noted that such a choice is only possible when the system has both
time-reversal symmetry and particle-hole symmetry. These two
symmetries also require $\Tau\Delta_\vk^\dagger$ to be Hermitian,
which makes the matrix $h_\vk+i\Tau\Delta_\vk^\dagger$ generically
non-Hermitian. The matrix $h_\vk+i\Tau\Delta_\vk^\dagger$ can be
decomposed by singular value decomposition (SVD) as
$h_\vk+i\Tau\Delta_\vk^\dagger=U_\vk^\dagger D_\vk V_\vk$ with
$U_\vk, V_\vk$ unitary matrices and $D_\vk$ a diagonal matrix with
non-negative elements. One can see straightforwardly that the
diagonal elements of $D_\vk$ are actually the positive eigenvalues
of $H_\vk$. For a fully gapped superconductor, $D_\vk$ is positive
definite, and we can adiabatically deform it to the identity
matrix $\mathbb{I}$ without closing the superconducting gap.
During this deformation the matrix $h_\vk+i\Tau\Delta^\dagger_\vk$
is deformed to a unitary matrix $Q_\vk=U_\vk^\dagger V_\vk\in {\rm
U(N)}$. As shown in Ref. \cite{schnyder2008}, the integer-valued
topological invariant characterizing topological superconductors
is defined as the winding number of $Q_\vk$:
\begin{eqnarray}
N_W=\frac1{24\pi^2}\int d^3{\bf k}\epsilon^{ijk}{\rm
Tr}\left[{Q^\dagger_{\bf k}\partial_iQ_{\bf k}Q^\dagger_{\bf
k}\partial_jQ_{\bf k}Q^\dagger_{\bf k}\partial_kQ_{\bf
k}}\right].\label{Windingnumber}
\end{eqnarray}

\begin{figure}[h]
\begin{center}
\includegraphics[width=3in] {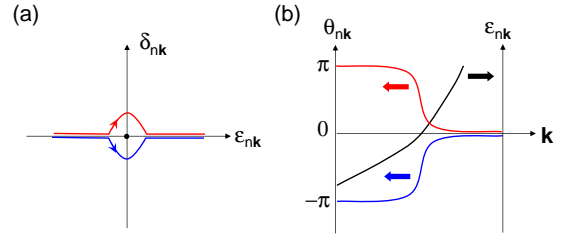}
\end{center}
\caption{(a) The path of $\epsilon_{n\vk}+i\delta_{n\vk}$ in the
complex plane for positive (red) and negative (blue)
$\delta_{n\vk}$ around the Fermi surface. (b) $\theta_{n\vk}$ and
$\epsilon_{n\vk}$ vs momentum $\vk$. The change of $\theta_{n\vk}$
across $k_F$ is $-\pi$ ($+\pi$) when $\delta_{n\vk}$ is positive
(negative), as shown by the red (blue) curve.} \label{fig:thetank}
\end{figure}

Now we study $Q_\vk$ in the weak pairing limit. For simplicity, in
the following, we will assume the Fermi surfaces are all
non-degenerate, and there are no lower dimensional zero-energy
defects such as point or line nodes. All our conclusions can be
easily generalized to more generic cases. When the Fermi surfaces
are non-degenerate, and the weak pairing term $\Delta_\vk$ is only
turned on around the Fermi surfaces, the matrix elements of
$\Tau\Delta_\vk^\dagger$ between different bands are negligible.
Thus, to  leading order we have
\begin{eqnarray}
h_\vk+i\Tau\Delta_\vk^\dagger&\simeq&
\sum_{n}\left(\epsilon_{n\vk}+i\delta_{n\vk}\right)\left|n,\vk\right\rangle\left\langle
n,\vk\right|\label{perturb}\\
\text{with~}\delta_{nk}&\equiv&\left\langle
n,\vk\right|\Tau\Delta_\vk^\dagger\left|n,\vk\right\rangle\in\mathbb{R}\nonumber
\end{eqnarray}
where $\left|n,\vk\right\rangle$ are the eigenvectors of $h_\vk$.
Physically, $\delta_{n\vk}$ is the matrix element of
$\Delta_{\vk}^\dagger$ between $\left|n,\vk\right\rangle$ and its
time-reversed partner
$\left|\bar{n},-\vk\right\rangle=\Tau^\dagger\left|n,\vk\right\rangle$.
In this approximation, the matrix $Q_\vk$ is given by
\begin{eqnarray}
Q_\vk=\sum_{n}e^{i\theta_{n\vk}}\left|n,\vk\right\rangle\left\langle
n,\vk\right|
\end{eqnarray}
with
$e^{i\theta_{n\vk}}=\left(\epsilon_{n\vk}+i\delta_{n\vk}\right)/\left|\epsilon_{n\vk}+i\delta_{n\vk}\right|$.
In the weak pairing limit, we take $\delta_{n\vk}$ to be nonzero
only in a small neighborhood $-\epsilon\leq E\leq \epsilon$ of the
Fermi level. As shown in Fig. \ref{fig:thetank}, the phase
$\theta_{n\vk}$ changes from $0$ to $\pm \pi$ across the Fermi
level, with the sign determined by the sign of $\delta_{n\vk}$. In
the limit $\epsilon\rightarrow 0$, such a domain wall
configuration of $\theta_{n\vk}$ can be expressed by the formula
\begin{eqnarray}
\nabla\theta_{n\vk}=-\pi{\bf v}_{n\vk}{\rm
sgn}\left(\delta_{n\vk}\right)\delta\left(\epsilon_{n\vk}\right)\label{thetadomain}
\end{eqnarray}
in which ${\bf v}_{n\vk}=\nabla_{\vk}\epsilon_{n\vk}$ is the Fermi
velocity. It should be noted that for a gapped superconductor
$\delta_{n\vk}$ remains nonzero for all $\vk$ on the Fermi
surfaces, so the sign of $\delta_{n\vk}$ is fixed on each Fermi
surface.

Once the behavior of $\theta_{n\vk}$ in the Brillouin zone is
simplified to Eq. (\ref{thetadomain}) in the weak pairing limit,
the winding number (\ref{Windingnumber}) can be simplified to the
following simple FSTI:
\begin{eqnarray}
N_W=\frac12\sum_{s}{\rm sgn}(\delta_s)C_{1s}\label{WindingFS}
\end{eqnarray}
where $s$ is summed over all disconnected Fermi surfaces, and
${\rm sgn}(\delta_s)$ denotes the sign of $\delta_{n\vk}$ on the
$s$-th Fermi surface. $C_{1s}$ is the first Chern number of the
$s$-th Fermi surface (denoted by ${\rm FS}_s$):
\begin{eqnarray}
C_{1s}=\frac1{2\pi}\int_{{\rm
FS}_s}d\Omega^{ij}\left(\partial_ia_{sj}({\bf
k})-\partial_ja_{si}({\bf k})\right)
\end{eqnarray}
with $a_{si}=-i\left\langle s\vk\right|\partial/\partial
k_i\left|s\vk\right\rangle$ the adiabatic connection defined for the
band $\left|s\vk\right\rangle$ which crosses the Fermi surface, and
$d\Omega^{ij}$ the surface element $2$-form of the Fermi surface.
More details of the derivation of Eq. (\ref{WindingFS}) are included
in the supplementary material\cite{SOM}.

As an example, consider a two-band Hamiltonian
$h_\vk={\vk^2}/{2m}-\mu+\alpha\vk\cdot\sigma$ For $\mu>0$, the
system has two Fermi surfaces which are concentric spheres around
$\vk=0$. (The two-band model should be regularized on the lattice,
but the lattice regularization is unimportant as long as no other
Fermi surfaces are introduced.) Denoting the electron states at
the inner (outer) Fermi surface by $\left|\vk,+(-)\right\rangle$,
we have $\sigma\cdot
\vk\left|\vk,\pm\right\rangle=\pm\left|\vk\right|\left|\vk,\pm\right\rangle$.
It is easy to check that the two Fermi surfaces carry opposite
Chern number $C_\pm=\pm 1$. Thus, according to (\ref{WindingFS}),
we can obtain a topological superconductor if the two Fermi
surfaces have opposite signs of pairing. The time-reversal matrix
is $\Tau=i\sigma_y$ in this system. If we have
$\Delta_\vk=i\Delta_0\sigma_y$, then
$i\Tau\Delta^\dagger_\vk=\Delta_0\mathbb{I}$ which has the same
sign on the two Fermi surfaces and leads to $N_W=0$. On the other
hand, if we have $\Delta_\vk=i\Delta_0\sigma_y\sigma\cdot\vk$,
then $i\Tau\Delta^\dagger_\vk=\Delta_0\sigma\cdot\vk$ has opposite
sign on the two Fermi surfaces, so that $N_W=1$ if $\Delta_0>0$.
If we take the limit $\alpha\rightarrow 0$, we obtain a
topological superconductor with quadratic kinetic energy term and
pairing $\Delta_\vk=i\Delta_0\sigma_y\sigma\cdot\vk$, which is
exactly the BdG Hamiltonian of the $^3{\rm He}$B phase. This
example also illustrates how the FSTI (\ref{WindingFS}) can be
generalized to systems with degeneracies on the Fermi surface: One
can always add a small perturbation proportional to
$\Tau\Delta_\vk^\dagger$ to the Hamiltonian to lift the
degeneracy, while preserving the topological properties of the
superconductor.

{\bf Dimensional reduction to 2D.} The FSTI  can be generalized to
lower dimensions \emph{i.e.} 2D and 1D. The TRI topological
superconductors in 2D and 1D are related to the one in 3D by {\it
dimensional reduction}, similar to the procedure carried out in
the context of TRI topological insulators in Ref. \cite{qi2008}.

\begin{figure}[h]
\begin{center}
\includegraphics[width=3in] {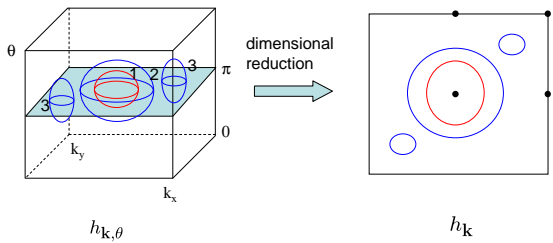}
\end{center}
\caption{Dimensional reduction from a 3D TRI superconductor to a 2D
TRI superconductor. The 2D TRI superconductor corresponds to the
$\theta=\pi$ section of a 3D superconductor. The Fermi surfaces with
blue (red) color are those with positive (negative) pairing
amplitude $\delta_s$. } \label{fig:dimred}
\end{figure}

Due to the same symmetry reason as the 3D case, the BdG
Hamiltonian $H_\vk$ of a 2D TRI superconductor can also be written
in the form of (\ref{HBdG}), so that one can also define a matrix
$Q_\vk\in{\rm U(N)}$ for the 2D case. Since $\Pi_2(U(N))=0$, we
can always find a smooth deformation
$Q_{\vk,\theta},~\theta\in[0,\pi]$ which interpolates between
$Q_\vk$ and the identity $\mathbb{I}$:
\begin{eqnarray}
Q_{\vk,\theta}=\left\{\begin{array}{cc}\mathbb{I},&\theta=0\\
Q_\vk,&\theta=\pi\end{array}\right.
\end{eqnarray}
It should be noted that $Q_\vk$ satisfies $\Tau^\dagger Q_\vk
\Tau=Q_{-\vk}^T$ due to time-reversal symmetry. Thus, if we define
$Q_{\vk,-\theta}=\Tau^\dagger Q_{-\vk,\theta}^T\Tau$ for
$\theta\in[0,\pi]$, we obtain $Q_{\vk,\theta}$ for
$\theta\in[-\pi,\pi]$ which is continuous and periodic in
$\theta\rightarrow \theta+2\pi$. Considering $\theta$ as a
momentum in an additional dimension, $Q_{\vk,\theta}$ describes a
3D TRI superconductor, which is characterized by the winding
number (\ref{Windingnumber}). If there are two different
interpolations $Q_{\vk,\theta}$ and $Q'_{\vk,\theta}$ which both
interpolate between $Q_{\vk}$ and $\mathbb{I}$, it can be shown
that time-reversal symmetry requires their winding numbers to be
different by an even number: $N_W(Q)-N_W(Q')=0~{\rm mod}~2$. Thus
the parity $(-1)^{N_W(Q)}$ is independent of the choice of
interpolation path, and is a $Z_2$ topological invariant uniquely
determined by $Q_\vk$.

Now we study the expression of such a $Z_2$ invariant in the weak
pairing limit. In this limit, the interpolation of $Q_\vk$ to
$Q_{\vk,\theta}$ is equivalent to interpolating the 1D Fermi
circles of the 2D normal state Hamiltonian $h_\vk$ to Fermi
surfaces in a 3D Brillouin zone parameterized by
$(k_x,k_y,\theta).$ We can simply extrapolate the pairing on the
Fermi circles to the Fermi surfaces, as illustrated in Fig.
(\ref{fig:dimred}). If the Fermi surfaces remain nondegenerate
during the interpolation, we obtain the $Z_2$ FSTI as the parity
of the winding number given by Eq. (\ref{WindingFS}):
\begin{eqnarray}
N_{2D}=(-1)^{N_W}=(-1)^{\frac12\sum_s{\rm
sgn}(\delta_s)C_{1s}}=\prod_{s}\left(i{\rm
sgn}(\delta_s)\right)^{C_{1s}}.\nonumber
\end{eqnarray}
Such a formula can be further simplified by noticing the following
two properties of the Chern number $C_{1s}$ carried by the Fermi
surfaces: i) The Chern number of each Fermi surface satisfies
$(-1)^{C_{1s}}=(-1)^{m_s}$, where $m_s$ is the number of TRI points
enclosed by the $s$-th Fermi surface. ii) The net Chern number of
all Fermi surfaces vanishes, $\sum_{s}C_{1s}=0$. We will leave a
more detailed demonstration of these two conclusions to the
supplementary material\cite{SOM}, and only sketch the physical
reasons for them here. The conclusion i) comes from the fact that a
Fermi surface which only encloses one TRI point, such as Fermi
surface 1 in Fig. \ref{fig:dimred}, always enclose a singularity at
the TRI point due to Kramers' degeneracy. One can prove that the
Chern number is always odd by making use of time-reversal symmetry.
The Fermi surfaces enclosing multiple TRI points can be
adiabatically deformed into several Fermi surfaces, each enclosing a
single TRI point. The conclusion ii) is a consequence of the
Nielsen-Ninomiya theorem\cite{nielson1981} which states that the
total chirality of a 3D lattice system must be zero.

Using these properties of $C_{1s}$, we finally obtain the
following expression for the $Z_2$ FSTI which is independent of
the interpolation to 3D:
\begin{eqnarray}
N_{2D}=\prod_{s}\left({\rm
sgn}(\delta_s)\right)^{m_s}.\label{Z22d}
\end{eqnarray}
The criterion shown in Eq. (\ref{Z22d}) is quite simple: a 2D TRI
superconductor is nontrivial (trivial) if there are an odd (even)
number of Fermi surfaces each of which encloses one TRI point in
the Brillouin zone and has negative pairing.

\begin{figure}[h]
\begin{center}
\includegraphics[width=3in] {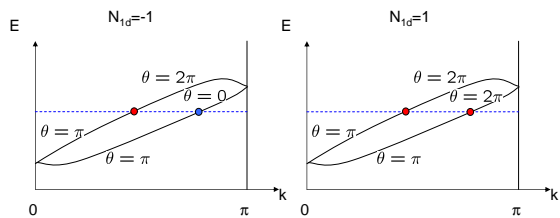}
\end{center}
\caption{Simple examples of (a) nontrivial and (b) trivial pairing
in a 1D system. The red and blue dots are Fermi points with
negative and positive pairing, respectively. The phase
$\theta=\theta_{n\vk}$ is taken to be the same at $k=0$ for the
two bands. At $k=\pi$, $\theta$ of the two bands are differ by
$2\pi$ for nontrivial pairing (a) and by $0$ for trivial pairing
(b).} \label{fig:1d}
\end{figure}

{\bf Dimensional reduction to 1D and generic expression of the
$Z_2$ invariant.} Following the same logic, the dimensional
reduction can be carried out again to obtain the $Z_2$ FSTI in 1D.
This results in an identical formula to Eq. (\ref{Z22d}). Since in
1D each Fermi ``surface" (which consists of two points at $k_F$
and $-k_F$) always encloses one TRI invariant point, the FSTI is
simply
\begin{eqnarray}
N_{1D}=\prod_s\left({\rm sgn}(\delta_s)\right)\label{Z21d}
\end{eqnarray}
where $s$ is summed over all the Fermi points between $0$ and
$\pi$. In other words, a 1D TRI superconductor is nontrivial
(trivial) if there are an odd number of Fermi points between $0$
and $\pi$ with negative pairing. Two examples with trivial and
nontrivial pairing are shown in Fig. \ref{fig:1d}.

Interestingly, from Fig. \ref{fig:1d} we can get an alternative
understanding of the 1D topological superconductor, which can
apply to a generic 1D TRI superconductor beyond the weak pairing
limit. As discussed earlier in Fig. (\ref{fig:thetank}), the sign
of the pairing $\delta_s$ determines the winding of the phase
$\theta_{n\vk}$ across the Fermi point. On the other hand, we have
shown that time-reversal symmetry requires $\Tau^\dagger Q_\vk
\Tau=Q_{-\vk}^T$, from which we can find that
$\theta_{n\vk}=\theta_{\bar{n}-\vk}$ if $\left|n,\vk\right\rangle$
and $\left|\bar{n},-\vk\right\rangle$ label a Kramers' pair. Thus,
along the path from $k=0$ to $k=\pi$, the change of
$\theta_{n\vk}$ and $\theta_{\bar{n}\vk}$ must be the same modulo
$2\pi$: $\int_0^\pi
dk\left(\partial_k\theta_{nk}-\partial_k\theta_{\bar{n}k}\right)=2\pi
n,~n\in\mathbb{Z}$. In the examples shown in Fig. \ref{fig:1d} we
have $n=1$ for the nontrivial pairing and $n=0$ for the trivial
pairing. Such a parity difference of the winding number of
$\theta_{n\vk}$ turns out to be generic, and can be captured by
the following $Z_2$ FSTI:
\begin{eqnarray}
 N_{1D}=\frac{{\rm Pf}\left({\Tau^\dagger Q_{k=\pi}}\right)}{{\rm
Pf}\left(\Tau^\dagger Q_{k=0}\right)}\exp\left(-\frac12\int_0^\pi dk
{\rm
Tr}\left[{Q_k^\dagger\partial_kQ_k}\right]\right)\label{Pfaffian}
\end{eqnarray}
where we have used $\Tau^\dagger Q_{k}\Tau=Q_{-k}^T\Rightarrow
\Tau^\dagger Q_{k}=-\left(\Tau^\dagger Q_{-k}\right)^T$, so that
$\Tau^\dagger Q_k$ is anti-symmetric for $k=0,\pi$, and the Pfaffian
is well-defined. It is straightforward to show that $N_{1D}=\pm 1$
is a $Z_2$ quantity, and also a topological invariant. More details
on the properties of the $Z_2$ FSTI (\ref{Pfaffian}) and its
relation to the FSTI (\ref{Z21d}) are given in the supplementary
materials.\cite{SOM} Eq. (\ref{Pfaffian}) is the topological
superconductor analog of Kane and Mele's $Z_2$ invariant in quantum
spin Hall insulators\cite{kane2005B}. Following the same approach as
Ref. \cite{fu2007b,moore2007}, one can obtain three $Z_2$ invariants
in 2D, one of which is the ``strong topological invariant"
$N_{2D}=N_{1D}(k_y=0)N_{1D}(k_y=\pi)$, with $N_{1D}(k_y=0(\pi))$ the
$1D$ topological invariant defined for the $k_y=0(\pi)$ system,
respectively. This topological invariant is robust to disorder, and
is equivalent to the one described by Eq. (\ref{Z22d}).

In summary, we have presented the criteria for TRI topological
superconductivity in the physical dimensions one, two and three.
When the Fermi surfaces are nondegenerate, the criteria are very
simple. In three dimensions, the winding number is an integer which
is determined by the sign of pairing order parameter and the Chern
number of the Fermi surfaces. In one and two dimensions, a pairing
around the Fermi surface is nontrivial if there are an odd number of
Fermi surfaces with a negative pairing order parameter. We also
obtained an explicit formula for the $Z_2$ invariants applicable to
generic 1D and 2D TRI superconductors. Our results provide simple
and physical criteria that can be used in the search of topological
superconductors.  Our FSTI's suggest to search for topological
superconductors in the nonconventional superconductors with strong
inversion symmetry breaking and strong correlation. The strong
inversion symmetry breaking is necessary to generate spin-split
Fermi surfaces, and strong electron-electron Coulomb interactions
prefer the pairing to have a nonuniform sign in the Brillouin
zone.\cite{scalapino1986}

{\bf Acknowledgement.}--We acknowledge N. Nagaosa for helpful
discussions. This work is supported by the US Department of Energy,
Office of Basic Energy Sciences under contract DE-AC03-76SF00515.

\begin{widetext}

\appendix

\section{Background and Derivation of the BdG Hamiltonian}
We first list some basic properties of time-reversal invariant
superconductors in generic dimensions and then go on to derive the
form of Eq. (1) from the Letter. Consider a general TRI
superconductor with the Hamiltonian
\begin{equation}
H=\sum_{\vk}\left[\psi_\vk^\dagger
h_\vk\psi_\vk+\frac12\left(\psi_{\vk}^\dagger\Delta_\vk\psi_{-\vk}^{\dagger
T}+H.c.\right)\right]\equiv\sum_\vk\kc{\psi_\vk^\dagger,\psi_{-\vk}^{T}}H(\vk)\vect{\psi_\vk\\\psi_{-\vk}^{\dagger
T}}\end{equation} with \bea H({\bf k})=\kc{\ba{cc}h({\bf
k})&\Delta({\bf k})\\\Delta^\dagger({\bf k})&-h^T(-{\bf k})\ea} \eea
The normal state Hamiltonian $h(\vk)$ is time-reversal invariant,
which means there is a matrix $\Tau$ satisfying \bea
\mathbb{T}^{-1}\psi_\vk
\mathbb{T}=\Tau^\dagger\psi_{-\vk},~\Tau^\dagger h_\vk
\Tau=h_{-\vk}^T,~\Tau=-\Tau^T,~\Tau^\dagger \Tau=\eye2 \eea From the
transformation property of $\psi_\vk$ we can obtain \bea
\mathbb{T}^{-1}\psi_{\vk}^\dagger \mathbb{T}=\psi_{-\vk}^\dagger
\Tau \eea so that \bea
\mathbb{T}^{-1}\vect{\psi_{\vk}\\\psi_{-\vk}^{\dagger T}}
\mathbb{T}\equiv\kc{\ba{cc}\Tau^\dagger&\\&\Tau^T\ea}\vect{\psi_{\vk}\\\psi_{-\vk}^{\dagger
T}}=
\kc{\ba{cc}\Tau^\dagger&\\&-\Tau\ea}\vect{\psi_{\vk}\\\psi_{-\vk}^{\dagger
T}}\eea and the time-reversal symmetry of the Hamiltonian requires
\bea T^\dagger
H(\vk)T=H(-\vk)^T,\text{~with~}T=\kc{\ba{cc}\Tau&\\&-\Tau^\dagger\ea}\label{T}
\eea On the other hand, the following identity \bea
\vect{\psi_{\vk}\\\psi_{-\vk}^{\dagger
T}}^\dagger=\kc{\ba{cc}&\eye2\\\eye2&\ea}\vect{\psi_{-\vk}\\\psi_{\vk}^{\dagger
T}} \eea requires the particle-hole symmetry of the BdG Hamiltonian:
\bea C^\dagger
H(\vk)C=-H(-\vk)^T,~\text{~with~}C=\kc{\ba{cc}&\eye2\\\eye2&\ea}\label{C}
\eea The two symmetries (\ref{T}) and (\ref{C}) require the pairing
matrix $\Delta(\vk)$ to satisfy \bea
\Delta(\vk)=-\Delta^T(-\vk),~\left(\Tau\Delta^\dagger(\vk)\right)^\dagger=\Tau\Delta^\dagger(\vk)
\eea If we define \bea
\chi=iTC^\dagger=\kc{\ba{cc}&i\Tau\\-i\Tau^\dagger&\ea}, \eea then
we have \bea \chi^\dagger H(\vk)\chi=CT^\dagger
H(\vk)TC^\dagger=CH(-\vk)^TC^\dagger=-H(\vk) \eea

The ``chirality operator" $\chi$ can be diagonalized by \bea
\chi=V^\dagger\kc{\ba{cc}\eye2&\\&-\eye2\ea}V,~\text{with}~V=\frac1{\sqrt{2}}\kc{\ba{cc}\eye2&-\eye2\\\eye2&\eye2\ea}\kc{\ba{cc}\eye2&\\&-i\Tau\ea}
\eea To derive Eq. 1 from the main letter we transform the basis to
the eigenbasis of $\chi$ to get the Hamiltonian form \bea
\tilde{H}({\bf k})=VH(\vk)V^\dagger=\kc{\ba{cc}&h_{\bf k}+i\Tau
\Delta_{\bf k}^\dagger\\h_{\bf k}-i\Tau\Delta_{\bf
k}^\dagger&\ea}\label{BdGHamiltonian} \eea As is mentioned in main
text, the matrix $A_{\bf k}=h_{\bf k}+i\Tau \Delta_{\bf k}^\dagger$
can be decomposed by singular value decomposition (SVD): \bea A_{\bf
k}&\equiv &h_{\bf k}+i\Tau \Delta_{\bf k}^\dagger=U^\dagger_{\bf
k}D_{\bf k}V_{\bf k} \eea in which $D_{\bf k}$ is a diagonal matrix
with nonnegative real diagonal components, and $U_{\bf k}$ and
$V_{\bf k}$ are unitary. The Hamiltonian $\tilde{H}({\bf k})$ can be
diagonalized as \bea
\tilde{H}({\bf k})&=&\kc{\ba{cc}&U^\dagger_{\bf k}D_{\bf k}V_{\bf k}\\V_{\bf k}^\dagger D_{\bf k}U_{\bf k}&\ea}\nn\\
&=&
\kc{\ba{cc}U^\dagger_{\bf k}&\\&V^\dagger_{\bf k}\ea}\kc{\ba{cc}&D_{\bf k}\\D_{\bf k}&\ea}\kc{\ba{cc}U_{\bf k}&\\&V_{\bf k}\ea}\nn\\
&=&\frac1{\sqrt{2}}\kc{\ba{cc}U_{\bf k}&U_{\bf k}\\-V^\dagger_{\bf
k}&V^\dagger_{\bf k}\ea} \kc{\ba{cc}D_{\bf k}&\\&-D_{\bf
k}\ea}\frac1{\sqrt{2}}\kc{\ba{cc}U^\dagger_{\bf k}&-V_{\bf
k}\\U^\dagger_{\bf k}&V_{\bf k}\ea} \eea Thus we see that the
eigenvalues of the Hamiltonian are given by the eigenvalues of
$D_{\bf k}$ and $-D_{\bf k}$. For a gapped Hamiltonian, all the
eigenvalues of $D_k$ are positive, so that we can adiabatically
deform ${D_{\bf k}}$ to $\eye2$, which deforms the Hamiltonian to
the form \bea \tilde{H}({\bf k})\backsimeq \kc{\ba{cc}&Q_{\bf
k}\\Q^\dagger_{\bf k}&\ea},~Q_{\bf k}\equiv U^\dagger_{\bf k}V_{\bf
k}\in {\rm U(N)} \eea It can be seen from the derivation above that
$Q_\vk$ is uniquely determined by the BdG Hamiltonian $H_\vk$, up to
a $\vk$-independent ${\rm U(N)\times U(N)}$ rotation \bea Q_\vk\>g
Q_\vk h,~g,h\in {\rm U(N)} \eea All physical information carried by
$Q_\vk$, such as the topological invariants, is insensitive to this
global ${\rm U(N)\times U(N)}$ rotation.

\section{Detailed Derivation of the 3d Fermi Surface Formula}
In this section we will show the detailed calculation of the $3d$
Fermi-surface formula. Beginning with the generic form of the
winding number in $3d$ we will show how to derive Eq. 6 from the
main letter. The general formula for the integer valued topological
number is \bea N_W=\frac1{24\pi^2}\int d^3{\bf
k}\epsilon^{ijk}\trace{Q^\dagger_{\bf k}\pa_iQ_{\bf k}Q^\dagger_{\bf
k}\pa_jQ_{\bf k}Q^\dagger_{\bf k}\pa_kQ_{\bf k}}. \eea

First of all, if $\Delta_{\bf k}=0$ for some region of ${\bf k}$,
the winding number density vanishes in that region. To see that,
notice that for $\Delta_\vk=0$, $Q_\vk$ is an adiabatic deformation
of $A_\vk=h_\vk$, so that $Q_\vk$ is Hermitian, and
$Q_\vk^2=Q_\vk^\dagger Q_\vk=\eye2$. Consequently the winding number
density is given by \bea
\rho_W=\frac1{24\pi^2}\epsilon^{ijk}\trace{Q_{\bf k}\pa_iQ_{\bf
k}Q_{\bf k}\pa_jQ_{\bf k}Q_{\bf k}\pa_kQ_{\bf k}}. \eea By making
use of $\pa_i\left(Q_\vk^2\right)=Q_\vk\pa_i Q_\vk+\pa_i Q_\vk
Q_\vk=0$, {\it i.e.}, $\ke{Q_\vk,\pa_i Q_\vk}=0$, one can prove that
$\rho_W=0$. This confirms our statement that in the weak pairing
limit, when only the pairing around Fermi surfaces is considered,
the topological invariant $N_W$ is completely determined by the
physics in the neighborhood of the Fermi surfaces.

As discussed in Eq. (4) of the letter, in the weak pairing limit
$Q_\vk$ can be written as
\begin{eqnarray}
Q_\vk=\sum_{n}e^{i\theta_{n\vk}}\left|n,\vk\right\rangle\left\langle
n,\vk\right|\label{spectrumofQ}
\end{eqnarray}
with
$e^{i\theta_{n\vk}}=\left(\epsilon_{n\vk}+i\delta_{n\vk}\right)/\left|\epsilon_{n\vk}+i\delta_{n\vk}\right|$
and $\delta_{nk}=\left\langle
n,\vk\right|\Tau\Delta_\vk^\dagger\left|n,\vk\right\rangle$. To the
leading order, near the Fermi surface we have \bea
e^{i\theta_{n\vk}}\simeq \frac{v_F\kc{k_\perp-k_F}+i\delta_{n
k_F}}{\sqrt{v_F^2\kc{k-k_F}^2+\delta_{n k_F}^2}} \eea In the limit
 $\delta_{nk_F}\>0$, we have \bea
\lim_{\delta_{nk_F}\>0}\theta_{n\vk}\>\pi{\rm
sgn}\kc{\delta_{nk_F}}\eta\kc{k_F-k_\perp} \eea with $\eta(x)$ the
step function satisfying $\eta(x)=1,~x\geq 0$ and $\eta(x)=0,~x<0$.
Thus we obtain
\begin{eqnarray}
\pa_{k_\perp}\theta_{n\vk}&=&-\pi{\rm
sgn}\kc{\delta_{n\vk}}\delta(k_\perp-k_F)\end{eqnarray} In the
vector form, this equation can be written as Eq. (5) of the letter:
\begin{eqnarray}
\nabla\theta_{n\vk}&=&-\pi{\bf v}_{n\vk}{\rm
sgn}\left(\delta_{n\vk}\right)\delta\left(\epsilon_{n\vk}\right)
\end{eqnarray}

Now we simplify the winding number formula by using Eq.
(\ref{spectrumofQ}). After some algebra we obtain
\begin{eqnarray}
N_W&=&\frac{i}{2\pi^2}\int d^3 k
\sum_{n,s}\epsilon^{ijk}\left[\partial_i\theta_n\left(a_{j}^{ns}\sin
\frac{\theta_{ns}}{2} \right)\left(a_{k}^{sn}\sin
\frac{\theta_{sn}}{2} \right)\right.\nonumber
\\ &-& \left.\frac{2i}{3}\sum_{p}\left(a^{pn}_{i}\sin
\frac{\theta_{pn}}{2}\right)\left(a^{ns}_{j}\sin
\frac{\theta_{ns}}{2}\right)\left(a^{sp}_{k}\sin
\frac{\theta_{sp}}{2}\right)\right]
\end{eqnarray}\noindent where $\theta_{ns}=\theta_n-\theta_s$ and $a^{ns}_{i}=-i\bra{n,{\bf k}}\pa_i\ket{s,{\bf k}}$ is the non-Abelian
adiabatic connection. When we restrict the pairing to an energy
shell $-\epsilon<\epsilon_{n\vk}<\epsilon$ and take the  $\epsilon
\> 0$ limit, the only nonvanishing term is the one with
$\partial_i\theta_n$ which has a $\delta$ function on the Fermi
surface. This leads to
\begin{eqnarray}
N_{W}&=&-\frac{i}{2\pi^2}\int_{FS} d^2{\bf
k_\parallel}\int_{k_F-\epsilon/v_F}^{k_F+\epsilon/v_F}dk_{\perp}
\sum_{n,s}\partial_{\perp}\theta_{n}\sin^{2}
\frac{\theta_{ns}}{2}\left(
a_{1}^{ns}a_{2}^{sn}-a_{2}^{ns}a_{1}^{sn}\right)\nn\\
&=&-\frac{i}{2\pi^2}\int_{FS} d^2{\bf k_\parallel}\sum_{n,s}\kd{\int
d\theta_\beta\sin^2\frac{\theta_n-\theta_s}2}\left(
a_{1}^{ns}a_{2}^{sn}-a_{2}^{ns}a_{1}^{sn}\right)\nn\\
&=&-\frac{i}{2\pi^2}\int_{FS} d^2{\bf
k_\parallel}\sum_{n,s}\left.\frac{\theta_n-\sin\kc{\theta_n-\theta_s}}2\right|_{\theta_n^-}^{\theta_n^+}\left(
a_{1}^{ns}a_{2}^{sn}-a_{2}^{ns}a_{1}^{sn}\right)
\end{eqnarray}\noindent
where $\theta_n^\pm$ are the values of $\theta_n$ right outside and
inside the Fermi surface, respectively. When there is only one band
that crosses the Fermi surface, $\theta_n^+=\theta_n^-$ for all
other bands. Labelling the single band crossing the Fermi surface
with $n=0$, we have \bea N_W&=&-\frac{i}{2\pi^2}\int_{FS} d^2{\bf
k_\parallel}\sum_{s\neq
0}\left.\frac{\theta_0-\sin\kc{\theta_0-\theta_s}}2\right|_{\theta_0^-}^{\theta_0^+}\left(
a_{1}^{0s}a_{2}^{s0}-a_{2}^{0s}a_{1}^{s0}\right). \eea Since
$\theta_0^{\pm}$ and $\theta_s$ all have the values $0$ or $\pi$,
the second term $\sin(\theta_0-\theta_s)$ vanishes, and we have \bea
N_W&=&-\frac{i}{4\pi^2}\int_{FS} d^2{\bf k_\parallel}\sum_{s\neq
0}\Delta\theta_0\left(
a_{1}^{0s}a_{2}^{s0}-a_{2}^{0s}a_{1}^{s0}\right)\nn\\
&=&\frac1{4\pi}\sum_{FS}\sgn\kc{\delta_{s\vk}}\int_{FS} d^2{\bf k_\parallel}\kc{\pa_1a_2^{00}-\pa_2a_1^{00}}\nn\\
&=&\frac12\sum_{s}\sgn\kc{{\delta_s}}C_{1s}.\label{FSformula3d} \eea
It should be noted that, (i) The superconducting gap on the Fermi
surface is given by $\abs{\delta_{n\vk}}$, so that
$\sgn{\kc{\delta_{n\vk}}}$ is the same for all the ${\bf k}$ on the
same Fermi surface, otherwise the superconducting gap would vanish
for some $\vk$. (ii) The Chern number of the $s$-th Fermi surface
$C_{1s}$ is defined with the normal vector along the direction of
${\bf v}_F$, which is opposite for an electron pocket and a hole
pocket.

\section{Proof of Chern-Number Properties Used in the Dimensional
Reduction to 2D section of Main Letter}

In the main letter, we have used the following two properties of the
Fermi surface Chern number to obtain the 2d $Z_2$ formula:
\begin{enumerate}
\item The Chern number of each Fermi surface satisfies
$(-1)^{C_{1s}}=(-1)^{m_s}$, where $m_s$ is the number of TRI points
enclosed by the $s$-th Fermi surface. \item The net Chern number of
all Fermi surfaces vanishes, $\sum_{s}C_{1s}=0$.
\end{enumerate}
In this section, we will prove both properties.

\subsection{Proof of Property 1}

We first study a simple Fermi surface enclosing one TRI point, {\it
e.g.}, the $\Gamma$ point, as shown in Fig. \ref{FS_supp} (a).
Denote the states at the Fermi level as $\ket{s,\vk}$. The Berry
phase gauge potential is defined by
$a^{ss}_{i}=-i\bra{s,\vk}\pa_i\ket{s,\vk}$. In the following, we
will denote $a_i=a^{ss}_i$ for simplicity. The time-reversal
invariance of the normal state Hamiltonian $h_\vk$ requires the
time-reversed state $T\kc{\ket{s,\vk}}$ to also be on the Fermi
surface. When the bands are non-degenerate on the Fermi surface, in
general we have \bea
T\kc{\ket{s,\vk}}&=&e^{i\varphi_\vk}\ket{s,-\vk}\label{Tofstates}\\
\=>a_i(-\vk)&=&-i\bra{s,-\vk}\frac{\pa}{\pa \kc{-k_i}}\ket{s,-\vk}=iT\kc{\bra{s,\vk}}e^{i\varphi_\vk}\pa_i\kd{e^{-i\varphi_\vk}T\kc{\ket{s,\vk}}}\nn\\
&=&\pa_i\varphi_\vk+i\kc{\bra{s,\vk}\pa_i{\kc{\ket{s,\vk}}}}^*\nn\\
&=&\pa_i\varphi_\vk+a_{i}(\vk).\label{Tofa} \eea Thus the gauge
curvature is \bea
f_{ij}(\vk)=\pa_ia_j(\vk)-\pa_ja_i(\vk)=-f_{ij}(-\vk). \eea We
denote the upper half of the Fermi surface with $k_z\geq0$ as ${\rm
FS}_+$ and the lower half as ${\rm FS}_-$. Thus \bea
C_{1s}=\frac1{2\pi}\int_{{\rm
FS}}d\Omega^{ij}f_{ij}(\vk)=\frac1{2\pi}\int_{{\rm
FS}_+}d\Omega^{ij}f_{ij}(\vk)+\frac1{2\pi}\int_{{\rm
FS}_-}d\Omega^{ij}f_{ij}(\vk) \eea Since the two form $d\Omega^{ij}$
denoting the normal direction of the Fermi surface is also odd in
$\vk$, the contributions of ${\rm FS}_+$ and ${\rm FS}_-$ to the
Chern number are equal, so that \bea C_{1s}=\frac1{\pi}\int_{{\rm
FS}_+}d\Omega^{ij}f_{ij}(\vk).\label{TRIC1} \eea Since ${\rm FS}_+$
is a manifold with boundary, the Chern form is equivalent to a
boundary integral: \bea C_{1s}=\frac1{\pi}\oint_{\pa {\rm
FS}_+}dl^ia_i(\vk)\label{polarization} \eea in which $\pa {\rm
FS}_+$ is the boundary of ${\rm FS}_+$, {\it i.e.}, the $k_z=0$
section of the Fermi surface, and $dl^i$ is the tangent vector to
$\pa {\rm FS}_+$. However, it should be noted that Eq.
(\ref{polarization}) holds only if $a_i(\vk)$ is continuous in the
whole ${\rm FS}_+$. In a generic gauge transformation
$a_i(\vk)\>a_i(\vk)+\pa_i\phi_\vk$
 on the boundary $\pa {\rm FS}_+$, the right hand side of Eq. (\ref{polarization}) can change by an even number:
 \bea
\frac1{\pi}\oint_{\pa {\rm FS}_+}dl^ia_i(\vk)\>\frac1{\pi}\oint_{\pa
{\rm FS}_+}dl^ia_i(\vk)+\frac1{\pi}\oint_{\pa {\rm
FS}_+}dl^i\pa_i\phi_\vk= \frac1{\pi}\oint_{\pa {\rm
FS}_+}dl^ia_i(\vk)+2n,~n\in\mathbb{N}
 \eea
Thus we have \bea C_{1s}=\frac1{\pi}\oint_{\pa {\rm
FS}_+}dl^ia_i(\vk)\mode{2}\label{polarization2} \eea in a generic
gauge choice.

Since the section $k_z=0$ of the Fermi surface is also symmetric
under time-reversal, we can split it to two parts $L_1$ and $L_2$,
which are the time reverse of each other, as shown in Fig.
\ref{FS_supp} (a). Noticing that the tangential vector $dl^i$ is
opposite for $\vk$ and $-\vk$, and by making use of Eq. (\ref{Tofa})
we have \bea
\oint_{L_1}dl^ia_i(\vk)&=&-\oint_{L_2}dl^ia_i(\vk)-\oint_{L_2}dl^i\pa_i\varphi_\vk\nn\\
\=>C_{1s}&=&\frac1\pi\oint_{\pa{\rm
FS}_+}dl^ia_i(\vk)=\frac1\pi\kc{\oint_{L_1}+\oint_{L_2}}dl^ia_i(\vk)=-\frac1\pi\oint_{L_2}dl^i\pa_i\varphi_\vk
.\eea One can always split the boundary so that there are only two
points $A$ and $B$ on the interface between $L_1$ and $L_2$. Due to
time-reversal symmetry, the two points must be the time-reversed
partners of each other, and the formula above becomes \bea
C_{1s}=-\frac1\pi\kc{\varphi_A-\varphi_B}\mode{2} \eea Denote the
momentum of $A$ and $B$ as $\vk_A$ and $\vk_B=-\vk_A$, according to
the definition Eq. (\ref{Tofstates}) we have \bea
T\kc{\ket{s,\vk_A}}&=&e^{i\varphi_A}\ket{s,\vk_B},~T\kc{\ket{s,\vk_B}}=e^{i\varphi_B}\ket{s,\vk_A}\nn\\
\=>T\kc{T\kc{\ket{s,\vk_A}}}&=&T\kc{e^{i\varphi_A}\ket{s,\vk_B}}=e^{-i\varphi_A}e^{i\varphi_B}\ket{s,\vk_A}
\eea On the other hand, we have $T^2=-1$ for each state, so that
\bea e^{i\kc{\varphi_A-\varphi_B}}=-1\=>C_{1s}=1\mode{2} \eea Thus
we have proved that $(-1)^{C_{1s}}=(-1)^{m_s}=-1$ for $m_s=1$.

\begin{figure}[h]
\begin{center}
\includegraphics[width=4in] {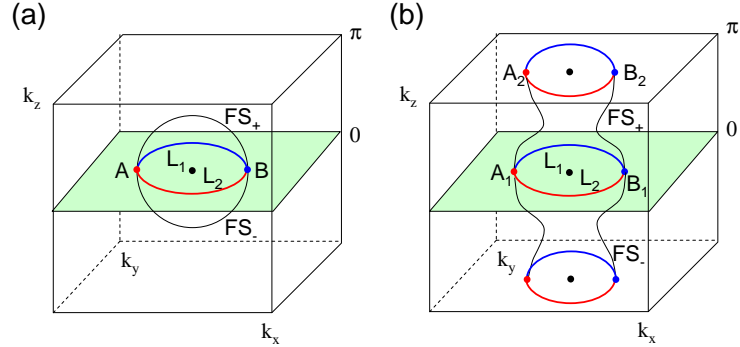}
\end{center}
\caption{(a) Schematic picture of a Fermi surface enclosing one TRI
point $(0,0,0)$. The Fermi surface is separated to two parts $FS_+$
and $FS_-$ by $k_z=0$ plane. The interface between the two parts is
further split into curves $L_1$ (red curve) and $L_2$ (blue curve)
which are time-reversal partner of each other. The interface of
$L_1$ and $L_2$ are given by points $A$ and $B$. (b) Schematic
picture of a Fermi surface enclosing two TRI points $(0,0,0)$ and
$(0,0,\pi)$. Similar to (a), the Fermi surface is separated to
$FS_+$ and $FS_-$, and the interface between $FS_+$ and $FS_-$ are
split into $L_1$ (red curve) and $L_2$ (blue curve), which intersect
at two pairs of points $A_{1}, B_1$ and $A_2,~B_2$. }
\label{FS_supp}
\end{figure}

For the Fermi surfaces enclosing more TRI points, as shown in Fig.
\ref{FS_supp} (b), the proof is similar. Due to the time-reversal
symmetry, we can always reduce the Chern number to an integral over
the upper half of the Fermi surface $FS_+$ as in Eq. (\ref{TRIC1})
and (\ref{polarization}). Generically, $FS_+$ has two boundaries at
$k_z=0$ and $k_z=\pi$, so that \bea C_{1s}=\frac1\pi\int_{\rm
FS_+}d\Omega^{ij}f_{ij}\kc{\vk}=\frac1\pi\kc{\oint_{\pa_\pi{\rm
FS}_+}-\oint_{\pa_0{\rm FS}_+}}dl^ia_i(\vk) \eea where
$\pa_{0,\pi}{\rm FS}_+$ stands for the boundary of ${\rm FS}_+$ at
$k_z=0$ and $k_z=\pi$ respectively. In the same way as above, the
boundary at $k_z=0$ can each be separated into two parts $L_1$ and
$L_2$, with several pairs of interface points
$A_i,B_i,~i=1,2,..,p_0$. By the same derivation as above one can
prove $\varphi_{A_i}-\varphi_{B_i}=\pi\mode{2\pi}$, and \bea
\frac1\pi{\oint_{\pa_{0}{\rm
FS}_+}}dl^ia_i(\vk)=-\frac1\pi\sum_{i=1}^p\kc{\varphi_{A_i}-\varphi_{B_i}}=p_0\mode{2}
\eea The same argument works for the $k_z=\pi$ boundary. Denoting
the number of interface points at the $k_z=\pi$ boundary by $p_\pi$,
we have \bea C_{1s}=p_\pi-p_0\mode{2}. \eea If the boundary
$\pa_{0,\pi}{\rm FS}_+$ encloses $m_{0,\pi}$ number of TRI points,
respectively, we have $p_0=m_0\mode2,~p_\pi=m_\pi\mode 2$. Thus \bea
(-1)^{C_{1s}}=(-1)^{p_\pi-p_0}=(-1)^{m_\pi+m_0}=(-1)^{m_s} \eea with
$m_s=m_\pi+m_0$ the total number of TRI points enclosed by the Fermi
surface. Thus we have proved the property 1.

\subsection{Proof of Property 2}

To prove property 2, we take a simple $s$-wave pairing \bea
\Delta_\vk=\Delta_0\Tau \eea with $\Delta_0$ a real number. For such
a pairing the matrix
$A_\vk=h_\vk+i\Tau\Delta_\vk^\dagger=h_\vk+i\Delta_0\eye2$, so that
the pairing on all the Fermi surfaces has the same sign: \bea
\delta_s=\bra{s,\vk}\Tau\Delta_\vk^\dagger\ket{s,\vk}=\Delta_0,~\forall
s .\eea Consequently, the topological invariant in the weak pairing
limit is given by \bea
N_W\kc{\Delta_0}=\frac12\sum_{s}\sgn\kc{\delta_s}C_{1s}=\frac{\sgn\kc{\Delta_0}}2\sum_sC_{1s}.\label{Nwforswave}
\eea

On the other hand, the BdG hamiltonian (\ref{BdGHamiltonian}) for
this simple pairing can be diagonalized easily to obtain the
eigenvalues \bea
E_{n\vk}^{\pm}=\pm\sqrt{\epsilon_{n\vk}^2+\Delta_0^2}. \eea Thus for
finite $\Delta_0$, the spectrum of the BdG Hamiltonian is always
gapped, so that the winding number $N_W(\Delta_0)$ remains invariant
for all $\Delta_0>0$. Thus we can compute $N_W$ in the limit
$\Delta_0\>+\infty$. The unitary matrix $Q_\vk$ is given by \bea
Q_\vk&=&\sum_{n}\ket{n,\vk}\frac{\epsilon_{n\vk}+i\Delta_0}{\sqrt{\epsilon_{n\vk}^2+\Delta_0^2}}\bra{n,\vk},\nn\\
\=>\lim_{\Delta_0\>+\infty}Q_\vk&=&i\sum_{n}\ket{n,\vk}\bra{n,\vk}=i\eye2
\eea Obviously, the winding number $N_W(\Delta_0\>+\infty)=0$, so
that $N_W(\Delta_0)=0$ for any $\Delta_0$. According to Eq.
(\ref{Nwforswave}) we have proven property 2: \bea \sum_sC_{1s}=0 .
\eea

\section{Properties of the 1d $Z_2$ Topological Invariant (11)}

In this section, we will study some basic properties of the $Z_2$
topological invariant defined in Eq. (11) of the letter, and show
how it is reduced to the Fermi surface formula (10) in the weak
pairing limit.

We start from Eq. (10) of the letter: \bea N_{1d}=\frac{{\rm
Pf}\kc{\Tau^\dagger Q_{k=\pi}}}{{\rm Pf}\kc{\Tau^\dagger
Q_{k=0}}}\exp\kc{-\frac12\int_0^\pi dk \trace{Q_k^\dagger\pa_kQ_k}}
\label{Pfaffian}\eea First of all, $\Tau^\dagger Q_{k}$ is
antisymmetric since \bea
\Tau^\dagger h_\vk \Tau=h_{-\vk}^T,~\Tau^\dagger \kc{\Tau\Delta_\vk^\dagger}\Tau=\Delta_\vk^\dagger\Tau=-\Delta_{-\vk}^{\dagger T}\Tau=\kc{\Tau\Delta_{-\vk}^\dagger}^T\nn\\
\=>\Tau^\dagger Q_{ k}\Tau=Q_{-k}^T\=>\Tau^\dagger
Q_{k}=Q_{-k}^T\Tau^\dagger=-\kc{\Tau^\dagger Q_{-k}}^T \eea Thus the
Pfaffian is well-defined at $k=0$ and $k=\pi$.

Since $Q_k\in {\rm U(N)}$, we have $\det{Q_k}=e^{i\varphi_k}$ which
is a ${\rm U(1)}$ phase. Since $\trace{Q_k^\dagger
\pa_kQ_k}=\trace{\log Q_k}=\log \det Q_k=i\varphi_k$, we have
$\int_0^\pi dk\trace{Q_k^\dagger
\pa_kQ_k}=i\kc{\varphi(\pi)-\varphi(0)}\mode{2\pi}$, so that \bea
\exp\kc{-\int_0^\pi dk
\trace{Q_k^\dagger\pa_kQ_k}}=e^{-i\kc{\varphi(\pi)-\varphi(0)}}=\frac{{\rm
det}\kc{\Tau^\dagger Q_{k=0}}}{{\rm det}\kc{\Tau^\dagger Q_{k=\pi}}}
\eea Thus \bea N_{1d}^2&=&\frac{{\rm det}\kc{\Tau^\dagger
Q_{k=\pi}}}{{\rm det}\kc{\Tau^\dagger Q_{k=0}}}\exp\kc{-\int_0^\pi
dk \trace{Q_k^\dagger\pa_kQ_k}}\equiv1 \eea so that $N_{1d}$ always
takes the value of $\pm 1$.

Now we show that $N_{1d}$ is a topological invariant. For an
infinitesimal deformation $Q_k\->Q_k'=Q_k+\delta Q_k$, the phase
factor $\exp\kc{-\frac12\int_0^\pi dk \trace{Q_k^\dagger\pa_kQ_k}}$
only depends on the deformation of $Q_k$ at $k=0$ and $\pi$: \bea
{\exp\kc{-\frac12\int_0^\pi dk
\trace{{Q'}_k^\dagger\pa_kQ'_k}}}=\exp\kc{-\frac12\int_0^\pi dk
\trace{Q_k^\dagger\pa_kQ_k}}e^{-\frac
i2\kc{\delta\varphi(\pi)-\delta\varphi(0)}} \eea On the other hand,
the change of Pfaffian is given by \bea {\rm Pf}\kc{\Tau^\dagger
Q'_{k=0,\pi}}=e^{\frac{i}2\delta\varphi_{k=0,\pi}}{\rm
Pf}\kc{\Tau^\dagger Q'_{k=0,\pi}} \eea Consequently, we see that
$\delta N_{1d}=0$ in any smooth deformation of the unitary matrix
$Q_\vk$ as long as time-reversal symmetry is preserved.

In the weak pairing limit, the general formula (\ref{Pfaffian}) can
be reduced to the Fermi surface formula given by Eq. (10) of the
letter. In the weak pairing limit, assume there are $M$ Fermi points
$k_{s},~s=1,2,...,M$ between $0$ and $\pi$. As discussed in the main
text, we require that the Fermi level does not cross any band at
$k=0$ or $\pi$. As discussed in Fig. 1 of the letter, each Fermi
point leads to a domain wall of $\theta_{s k}$ for the corresponding
band $s$ crossing the Fermi level. According to Eq.
(\ref{spectrumofQ}) in the weak pairing limit we have \bea
\det{Q_k}=\exp\kc{i\sum_{n}\theta_{nk}} \eea Across each Fermi point
$k_{Fs}$, the phase $\theta_{sk}$ will jump by
$-\pi\sgn\kc{v_{Fs}\delta_{sk_{s}}}$ and the $\theta_{nk}$ for other
bands remain invariant. It should be noted that the sign of $v_F$
enters the expression since the winding of $\theta_{sk}$ is given by
$-\pi\sgn\kc{\delta_{sk_{s}}}$ {\em along the direction of the Fermi
velocity $v_{Fs}$}. Consequently, the phase $\log \det
Q_k=i\sum_{n}\theta_{nk}$ is changed by
$-i\pi\sgn\kc{v_{Fs}\delta_{sk_{s}}}$ across the $s$-th Fermi point,
and the net change of $\log\det Q_k$ from $0$ to $\pi$ is given by
\bea
\int_0^\pi dk \pa_k\log\det Q_k&=&-i\pi\sum_{s=1}^M\sgn\kc{v_F\delta_{sk_{s}}}\nn\\
\=>\exp\kc{-\frac12\int_0^\pi dk
\trace{Q_k^\dagger\pa_kQ_k}}&=&\prod_{s}e^{-\frac{i\pi}2\sgn\kc{v_{Fs}\delta_{sk_{s}}}}\equiv \prod_s\kc{-i\sgn\kc{v_{Fs}\delta_{sk_{s}}}}\nn\\
&=&\prod_s\kc{\sgn\kc{\delta_{sk_s}}}\prod_s\kc{-i\sgn\kc{v_{Fs}}}\label{winding1d}
\eea When there are $m$ Fermi points with positive $v_{Fs}$ and $n$
Fermi points with negative $v_{Fs}$, $n-m$ gives the number of bands
which are above the Fermi level at $k=0$, but below the Fermi level
at $k=\pi$. If we denote $N_2(0)$ and $N_2(\pi)$ as the number of
bands occupied at $k=0$ and $k=\pi$, respectively, then
$n-m=N_2(\pi)-N_2(0)$. Since all bands are paired in Kramers pairs
at $k=0,\pi$, $N_2(0)$ and $N_2(\pi)$ must be even. Thus we have
\bea
\prod_s\kc{-i\sgn\kc{v_{Fs}}}=(-i)^{m}i^n=e^{\frac{i}2\pi\kc{n-m}}=(-1)^{\frac{N_2(\pi)-N_2(0)}2}\label{prodvf}
\eea

Now we study the Pfaffian ${\rm Pf}\kc{\Tau^\dagger Q_{k=0,\pi}}$.
Since we have assumed the Fermi level does not cross the bands at
$k=0,\pi$, in the weak pairing limit we have $\Delta_{k=0,\pi}=0$.
If the normal state Hamiltonian $h_k$ is diagonalized to \bea
h_k=U_k^\dagger\kc{\ba{ccc}\epsilon_1(k)&&\\&...&\\&&\epsilon_N(k)\ea}U_k
\eea $Q_k$ can be obtained by \bea
Q_k=U_k^\dagger\kc{\ba{cc}\eye2_{N_1\times N_1}&\\&-\eye2_{N_2\times
N_2}\ea}U_k \eea in which $N_1$ and $N_2$ are the number of
unoccupied and occupied bands, respectively. The Pfaffian ${\rm
Pf}\kc{\Tau^\dagger Q_k}$ can be obtained as \bea
{\rm Pf}\kc{\Tau^\dagger Q_k}&=&{\rm Pf}\kc{\Tau^\dagger U_k^\dagger\kc{\ba{cc}\eye2_{N_1\times N_1}&\\&-\eye2_{N_2\times N_2}\ea}U_k}\nn\\
&=&{\rm Pf}\kc{U_k^{\dagger T}\Tau^\dagger
U_k^\dagger\kc{\ba{cc}\eye2_{N_1\times N_1}&\\&-\eye2_{N_2\times
N_2}\ea}}\cdot\det{U_k} \eea By making use of the time-reversal
invariance condition $\Tau^\dagger Q_k\Tau=Q_{-k}^T$, one can prove
that \bea \kd{U_k^{\dagger T}\Tau^\dagger
U_k^\dagger,\kc{\ba{cc}\eye2_{N_1\times N_1}&\\&-\eye2_{N_2\times
N_2}\ea}}=0 \eea for $k=0,\pi$, so that the matrix $U_k^{\dagger
T}\Tau^\dagger U_k^\dagger$ is block diagonal. Consequently, we have
\bea {\rm Pf}\kc{U_k^{\dagger T}\Tau^\dagger
U_k^\dagger\kc{\ba{cc}\eye2_{N_1\times N_1}&\\&-\eye2_{N_2\times
N_2}\ea}}=(-1)^{N_2/2}{\rm Pf} \kc{U_k^{\dagger T}\Tau^\dagger
U_k^\dagger}=(-1)^{N_2/2}{\rm
Pf}\kc{\Tau^\dagger}\cdot\det{U_k^\dagger} \eea Thus \bea {\rm
Pf}\kc{\Tau^\dagger Q_k}&=&(-1)^{N_2/2}{\rm
Pf}\kc{\Tau^\dagger}\label{Pfweakpairing} \eea for $k=0,\pi$. It
should be noted that the number of occupied bands $N_2$ is always
even for $k=0,\pi$ due to Kramers degeneracy.

Combining Eq. (\ref{winding1d}), (\ref{prodvf}) and
(\ref{Pfweakpairing}) we obtain \bea
N_{1d}=\prod_s\kc{\sgn\kc{\delta_{sk_s}}} \eea Thus we have proved
that the general $Z_2$ invariant (11) in the main text is equivalent
to Eq. (10) in the main text in the weak pairing limit.

\end{widetext}

\end{document}